# Robust Super-Resolution Imaging Based on a Ring Core Fiber with Orbital Angular Momentum

## Author Information

### Affiliations

Zheyu Wu, Ran Gao*, Sitong Zhou, Fei Wang, Zhipei Li, Huan Chang, Dong Guo, Xiangjun Xin

**School of Information and electronics, Beijing Institute of Technology, Beijing, China**

Qi Zhang, Feng Tian

**School of Electronic engineering, Beijing University of Posts and Telecommunications, Beijing, China**

Qiang Wu

**Faculty of Engineering and Environment, Northumbria University, Newcastle Upon Tyne, United Kingdom**

Corresponding author's E-mail: *6120190142@bit.edu.cn*

### Contributions

Z.W. prepared the experimental setup and performed the experiments. S.Z. and F.W. implemented and tested the neural network. All authors discussed the results and contributed to the writing of the paper. R.G. conceived and led the project.

# Abstract

Single fiber imaging technology offers unique insights for research and inspection in difficult to reach and narrow spaces. In particular, ultra-compact multimode fiber (MMF) imaging, has received increasing interest over the past decade. However, MMF imaging will be seriously distorted when subjected to dynamic perturbations due to time-varying mode coupling, and the imaging of space objects via Gaussian beam will be relatively degraded at the edge due to insufficient contrast. Here, a robust super-resolution imaging method based on a ring core fiber (RCF) with orbital angular momentum (OAM) has been proposed and experimentally demonstrated. The OAM modes propagating in the RCF form a series of weakly-coupled mode groups, making our imaging system robust to external perturbations. In addition, a spiral phase plate is used as a vortex filter to produce OAM for edge enhancement, thus improving the image resolution. Furthermore, a few-shot U-Transformer neural network is proposed to enhance the resilience of the developed RCF-OAM imaging system against environmental perturbations. Finally, the developed RCF-OAM imaging system achieves biological image transmission, demonstrating the practicality of our scheme. This pioneering RCF OAM imaging system may have broad applications, potentially revolutionising fields such as biological imaging and industrial non-destructive testing.

# Introduction

Single fiber imaging technology (SFIT) is a cutting-edge method in the field of micro-endoscopes for capturing and transmitting images through optical fibers, offering a number of advantages, including a small footprint, high flexibility, and low propagation loss. Indeed, SFIT is widely used in the fields of biological endoscopy and industrial non-destructive testing to achieve a precise image insight into narrow spaces and hard-to-reach areas. In bioendoscopy, the small footprint and inert glass components of SFIT enable the non-invasive imaging of blood vessels, the deep brain, or other biological tissues [1-5].

Additionally, owing to the electromagnetic interference and corrosion resistance properties of optical fibers, SFIT plays a significant role in detecting cracks and corrosion in buildings or components [6].

In general, there are three main types of conventional SFIT. The first type is an optical fiber coherence tomography (OCT) technology, which is based on low-coherence interferometry for obtaining cross-sectional images of multi-layered samples. In 2011, Ishida *et al*. utilized a femtosecond ultrashort pulsed Er-doped fiber laser as the seed source to improve the axial resolution of OCT imaging [7]. Pahlevaninezhad *et al*. developed a new type of OCT fiber probe using a nanostructured metasurface to achieve near-diffraction-limited focusing resolution and high depth of focus [8]. However, OCT technology requires additional precisely controlled coherent light as a reference, which increases the complexity of the system [9, 10]. The second type is the fiber confocal microscope (FCM), which focuses the beam to the same diffraction-limited spot through an optical fiber and scans the sample in depth to produce an image of the internal structure of the sample. Ahn *et al.* combined FCM and a spectral-domain interferometer to simultaneously measure the diameter and depth of through silicon vias [11]. Gu *et al*. applied a stimulated emission depletion method based on a circular polarization polarized vortex beam in a two-photo FCM to improve the imaging resolution [12]. The FCM technology has higher spatial resolution than OCT. However, the former suffers from excessive photobleaching and requires an additional scanning system at the proximal or distal end [13]. The last type is the multi-core fiber (MCF) imaging technology, which utilizes a MCF bundle consisting of hundreds or thousands of single mode fiber cores, where each core could collect and transmit images individually, acting as image pixels. Combined with computational imaging methods, MCF imaging technology can achieve a high imaging resolution and flexibility [14-16]. Choi *et al.* reported a fully flexible ultrathin fiber endoscope obtaining 3D holographic images of unstained tissues [15]. Orth *et al*. proved that MCF can record depth information and realize 3D imaging [17]. However, MCF-based endoscopes have limitations such as low resolution, large footprint, and pixelated images.

Multimode fiber (MMF) imaging technology has attracted increasing interest in the past decade due to its capacity to support numerous orthogonal spatial modes for image transmission. Due to the modal dispersion caused by the difference in propagation velocity, thousands of propagated modes interfere and combine in superposition at the distal end of the MMF, forming seemingly random speckle patterns, which can be recovered through the transmission matrix (TM) or deep learning methods [18, 19]. Compared to conventional SFIT, MMF imaging technology offers unique advantages such as a low cost, compactness, and elimination of fluorescence operation. Moreover, the MMF imaging approach offers a footprint reduction exceeding an order of magnitude, indicating that it represents a crucial breakthrough in the field of microendoscopy.

Nonetheless, the main hurdles faced by MMF imaging technology are less robust to strong time-varying mode coupling and low imaging resolution. The mode coupling in MMF is highly sensitive to environmental perturbations, such as macro/micro bending or temperature fluctuations, resulting in significant changes in the TM of the optical field between the input and output of the MMF. This shift leads to severe degradation of the performance of the TM in perturbed environments, thereby directly affecting the imaging quality and causing image distortion. Various studies have been conducted to address this limitation. For example, to compensate for the bending effects of the MMF, a partial reflector from the MMF distal end was proposed to generate the feedback light as the coherent beacon source [20, 21]. Resisi *et al*. trained a convolutional neural network (CNN) in a large amount of data to learn the invariant properties of signals under different optical fiber configurations [22]. However, these methods involve multiple operational processes, which significantly increase the complexity of the imaging system. On the other hand, insufficient contrast in practical imaging results in relative degradation of bright-field imaging at the edges of spatial objects, thereby further affecting the imaging resolution. While the resolution of MMF imaging is restricted by the numerical aperture of the MMF [23]. The MMF imaging technology must thus be improved in order to satisfy the requirements of high resolution in the field of biological processes such as cellular dynamics and neural activities. Therefore, developing a robust and high resolution SFIT is crucial in the field of microendoscopy, especially in the fields of biological and neural research.

Here, we propose and experimentally demonstrate a robust super-resolution imaging method based on a ring core fiber (RCF) with orbital angular momentum (OAM). The OAM modes propagating in RCF formed: 1) a series of weakly-coupled inter-mode groups (MGs), and 2) strong-coupled intra-MGs. The strong-coupled OAM modes in the intra-MGs formed the speckle pattern in the RCF, featuring effective image transmission. Furthermore, the weak coupling between different inter-MGs could ensure that the transmitted information was not easily distorted due to environmental perturbations, making the RCF OAM image system more robust to small environmental disturbances. A few-shot U-Transformer neural network (FSUT-NN) is proposed to enhance the resilience of the developed RCF-OAM imaging system against environmental perturbations. A spiral phase plate (SPP) was used as a vortex filter to produce OAM, which could enhance the edge sharpness and thus improve the image resolution. The proposed RCF OAM image system has potential applications in numerous fields such as biological imaging and industrial non-destructive testing, etc.

## Results

Principles

**Fig. 1: Imaging process.** Schematic diagram of RCF OAM imaging system with neural network inversion algorithm.

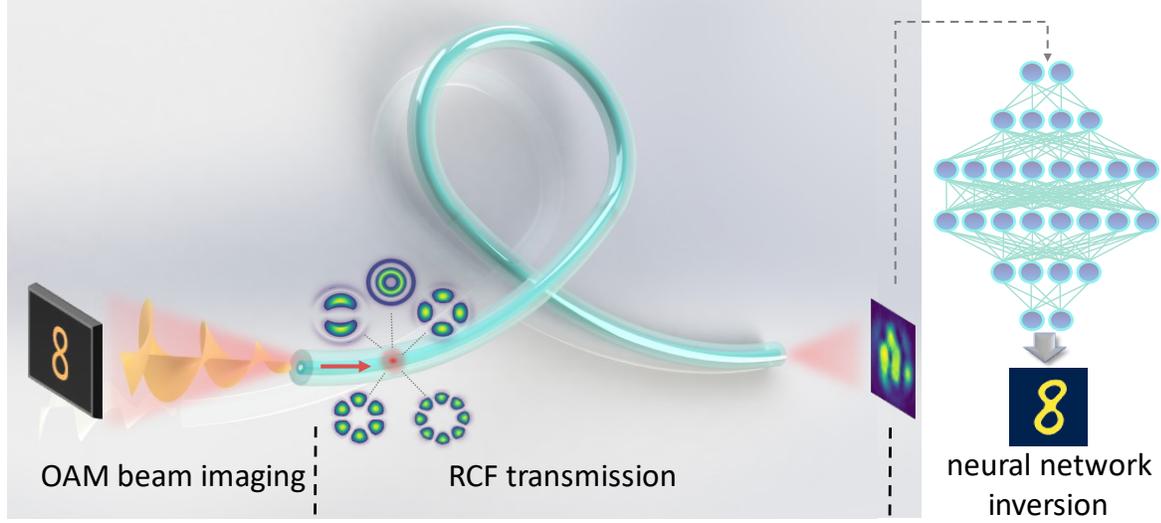

Fig. 1 presents a schematic diagram of the proposed RCF OAM system. The overall system consists of three parts, namely the OAM beam imaging, RCF transmission, and neural network inversion algorithm. First, in the OAM beam imaging, the light field of input images was converted into OAM form. In our experiment, a SPP with an azimuthal structure of $\exp(\pm il\varphi)$ was used to generate the OAM light beam, which was placed in the Fourier plane of a $4f$ system to implement a radial Hilbert transform to the propagated light beam. The transmittance function $H_l(\rho,\varphi)$ of the radial Hilbert transform can be express as [24]:

$$H_l(\rho,\varphi) = c\left(\frac{\rho}{R_0}\right)\exp(il\varphi) \qquad (1)$$

where $l$ is the order of SPP, $\rho$ and $\varphi$ are the polar coordinates on the plane of SPP, and $c\left(\frac{\rho}{R_0}\right)$ describes the circular aperture function with a radius of $R_0$. This method actually introduces OAM to the frequency spectrum of the object field at the filtering plane, which is commonly known as vortex filtering [25]. For a laterally extended planar object with an input function of $g(r,\varphi)$, the output image $\tilde{g}(r,\varphi)$ is formed by a convolution of the input function with the kernel function, which is also named the point spread function (PSF). The kernel function $h_l(r,\varphi)$ could be calculated using the Fourier transform of the filter function $H_l(\rho,\varphi)$ as:

$$h_l(r,\varphi) = -\frac{\pi R_0}{2r}\exp(il\varphi)\left[J_1(\tau)H_0(\tau) - J_0(\tau)H_1(\tau)\right] \qquad (2)$$

where $J_0$ and $J_1$ are Bessel functions of zero and first orders, and $H_0$ and $H_1$ are Struve functions of the zero and first orders, respectively. The parameter $\tau$ is given by $\tau = kR_0 r/f$, where $k$ is the wave vector of the light, and $f$ is the focal length of the Fourier transform lens. The profile of the kernel function from equation (2) is depicted in Fig. 2. The amplitude distribution of the kernel function is radial symmetric with respect to the zero-intensity centre, while the phase difference of the central symmetric position is $\pi$, as shown in Fig. 2 (a) and (b). Therefore, a signal was generated only when the system response of the input image was not perfectly offset in the direction of the gradient at the magnitude or phase gradient position, thereby achieving an edge enhancement, as shown in Fig. 2(c). In traditional MMF based imaging, the edge information of the image is easily affected by noise when a Gaussian beam is used, making it difficult to distinguish the edge information of the obtained image. In contrast, the OAM imaging scheme with vortex filtering emphasises the gradient information in the image, resulting in strong edge information, moreover enabling a high-contrast, isotropic edge enhancement, and high resolution optical fiber imaging.

**Fig. 2: Edge Enhancement Principle. a** Amplitude distribution of the kernel function of a Hilbert transform in polar coordinates. **b** Phase distribution of the kernel function of a Hilbert transform in polar coordinates. **c** Schematic diagram of filtering objects to achieve edge enhancement.

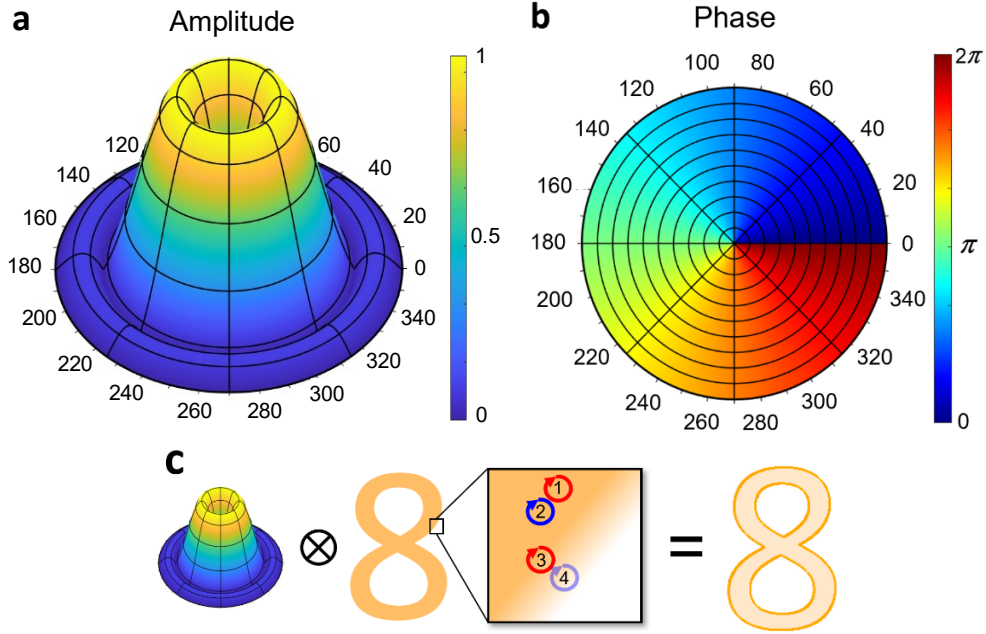

After the OAM light imaging, the original optical field in space could be directly coupled into the optical fiber, and multiple modes were excited in the optical fiber; the original information was thereby transmitted in the fiber through diverse mode distributions. Traditional fiber image transmission usually uses graded-index multimode fibers (GI-MMFs), whose cores have a parabolic refractive index distribution [26]. In GI-MMF, the image is coupled into many linearly polarized (LP) modes. According to the coupled mode theory, the optical power of different modes with different propagation constants ($\beta$) are coupled to each other. The coupling coefficient between two modes is greatly affected by the propagation constant difference ($\Delta\beta$), proportional to $1/\Delta\beta^4$ [27]. Meanwhile, $\Delta\beta$ is proportional to the effective refractive index difference ($\Delta n_{eff}$), i.e., $\Delta\beta = 2\pi\Delta n_{eff}/\lambda$. Therefore, the smaller the $\Delta n_{eff}$, the larger the coupling coefficient. When the fiber is disturbed, mode coupling between modes leads to the severe distortion of the information transmitted in the fiber. In GI-MMF, propagating LP modes with a similar effective refractive index are degenerated to an MG. As shown in Fig. 3**a**, there is a fixed $\Delta n_{eff}$ between different MGs, and the number of degenerate LP modes increases with the MG order. The large amount of degenerate LP modes of GI-MMF exhibits strong random mode coupling during transmission, as depicted in Fig. 3**b**. Hence, the stability of signal transmission in the GI-MMF is very weak due to the strong intra-MG coupling between a large number of degenerate LP modes [28], causing strong external interference to the GI-MMF fiber imaging system.

**Fig. 3: Mode transmission of GI-MMFs. a** Calculated effective index of supported modes in GI-MMF with a core diameter of 50 μm. The maximum relative index difference between the fiber core and cladding is 1%. **b** A schematic diagram of the changes to LP modes when GI-MMF are used for imaging.

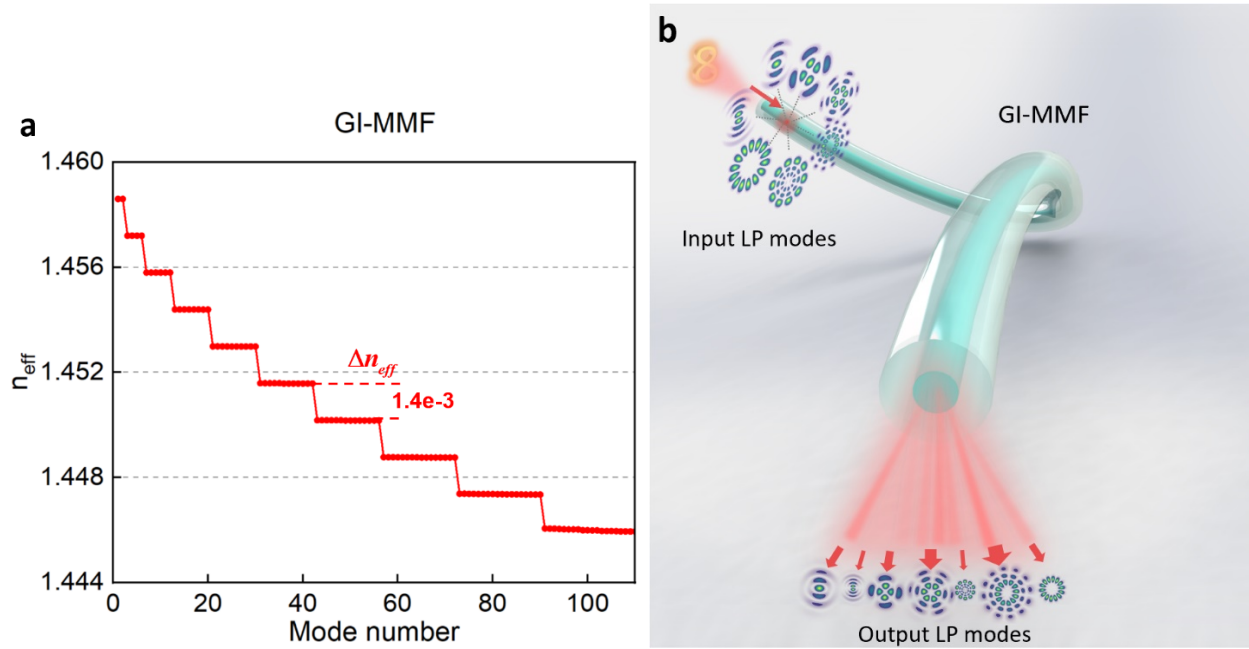

Ring core fibers have a unique structure and their core is not a solid circle but a ring with high refractive index. The inner core of the ring usually has a refractive index similar to that of cladding, thus forming an annual refractive index distribution, which has been proven to be a preferred design for the stable propagation of OAM modes [29]. In RCF, OAM modes with a similar refractive index is also defined as a MG. Therefore, four OAM modes (right circular (RC) with a polarization of +l, RC with a polarization of -l, left circular (LC) with a polarization of +l, and LC with a polarization of -l) with a topological charge of $\pm l$ will form in one MG, except that the first MG with a topological charge of 0 will contain two degenerate modes, as shown in Fig. 4**a**. Furthermore, compared to the relatively small $\Delta n_{eff}$ between adjacent MGs in GI-MMF, the $\Delta n_{eff}$ between adjacent MGs in RCF are much larger, making the inter-MG coupling in RCF weaker than that of GI-MMF [30]. The nature of the mode distribution in RCF leads to strong coupling within MG and weak coupling between MGs. Fig. **4b** depicts the OAM modes of the RCF used for imaging, showing weak random mode coupling throughout propagation. Consequently, the mode coupling in the RCF mainly occurred in one intra-MG with four modes, which was much less than the number of coupled modes in GI-MMF. Although the intra-MG coupling was also affected by external perturbations, the weak inter-MG made the RCF very robust to external disturbances. Therefore, the imaging transmission through the RCF was robust due to the strong intra-MG coupling of only four modes.

**Fig. 4: Mode transmission of RCFs. a** Calculated effective index of supported modes in a five MG RCF with a maximum relative index difference between the fiber core and cladding of 1%. **b** Schematic diagram of the changes to the OAM modes when RCF are used for imaging.

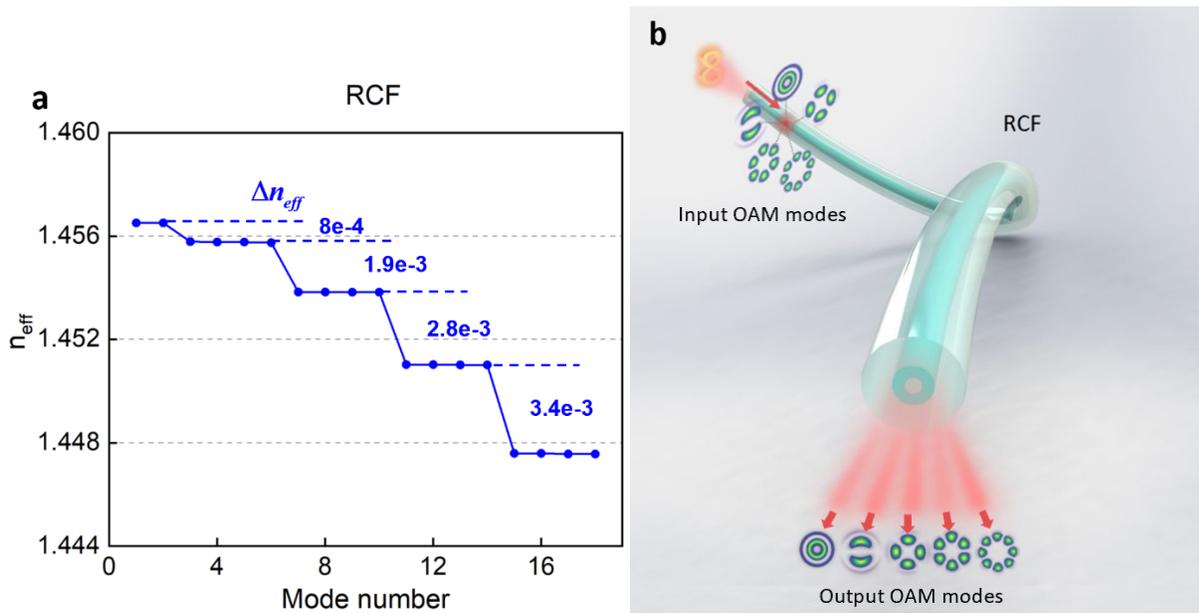

## Few-shot deep neural network

**Fig. 5: Neural network architecture. a** Schematic diagram of the proposed FSUT-NN architecture, which takes the general encoder-decoder structure. **b** Schematic diagram of the Swin-transformer block. **c** Schematic diagram of the Skip-layer excitation module.

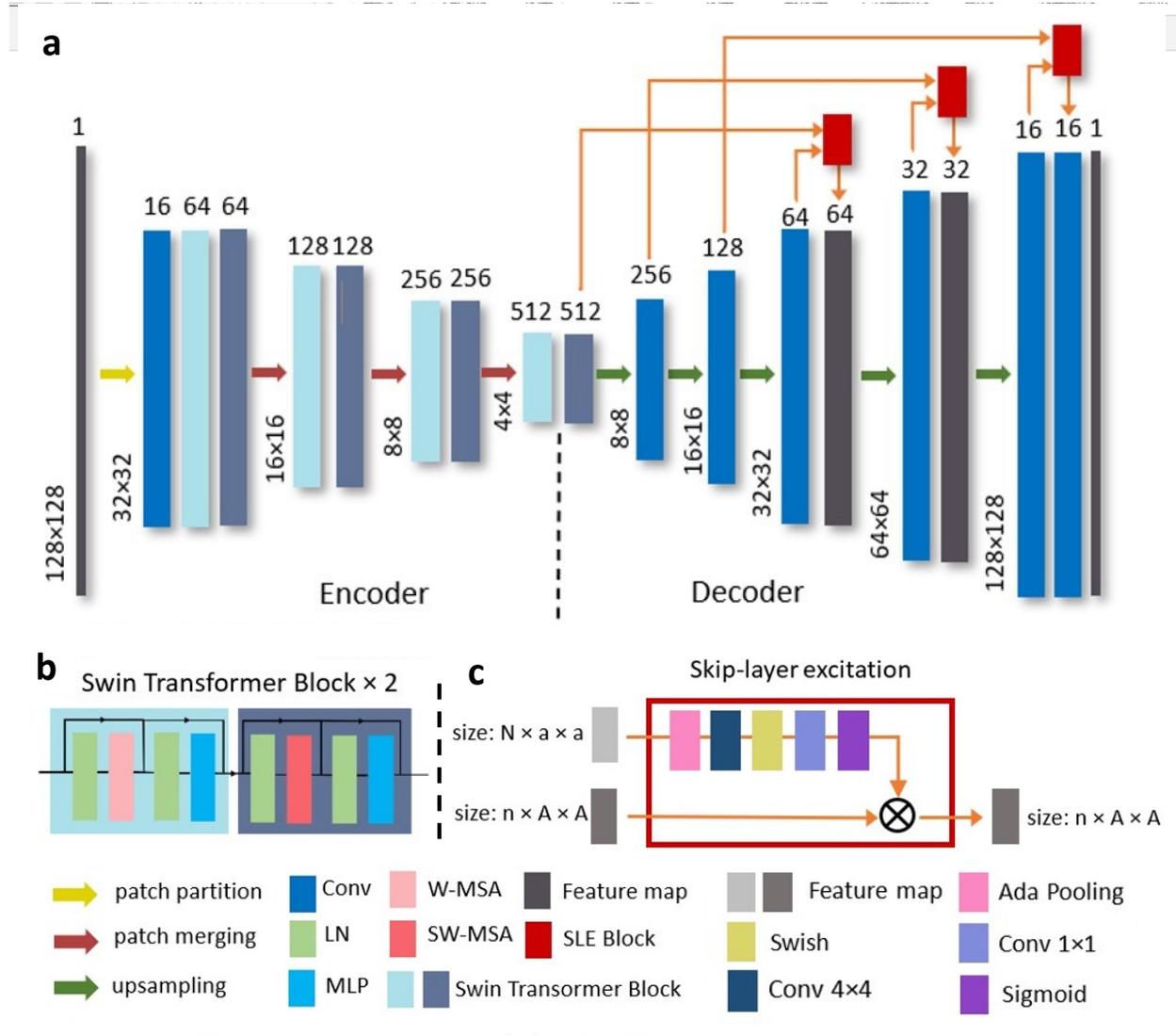

After the propagation in the RCF, the speckle pattern could be reconstructed to the original image using the neural network. Typically, neural networks follow an encoder-decoder architecture, such as a U-Net network, which extracts spackle patterns into a high-dimensional feature space through an encoder, and then recovers the features in high-dimensional space to the original images by a decoder. In our scheme, although the weak coupling between MGs in the RCF could effectively mitigate the crosstalk of external perturbations caused by random mode coupling, the strong coupling within the MGs was still affected by external perturbations under severe disturbances. To address this issue, a neural network with a strong generalization capability was needed. In this study, we chose a few-shot neural network to recover the speckle patterns in the RCF to the original image. In RCF, the speckle patterns may vary with external perturbations due to random mode coupling. Therefore, a training speckle pattern can be regarded as a "small sample", while the test speckle pattern can be considered a "large target". By treating the training speckle pattern as a finite sample, the few-shot neural network could enhance the generalization ability of the recovery algorithm. As a result, although the speckle pattern may change due to external perturbation, the test speckle pattern can be still recovered accurately by using the generalization ability of the few-shot neural network, which significantly improves the robustness of RCF imaging. Fig. 5**a** provides a schematic diagram of the few-shot deep neural network used for the image reconstruction, where an input image size of 128×128 pixels was assumed for all architectures. The "encoder" path was a typical Swin-Transformer architecture [31]. First, the input image was segmented into 4×4-sized non-overlapping blocks in the block segmentation layer. Here, the dimensionality of feature map was adjusted to $C = 64$ using a convolutional layer. Then, the feature map was downsampled and dimensionally augmented by four basic layers consisting of a pair of Swin transformer blocks. As shown in Fig. 5**b**, a pair of consecutive Swin transformer blocks contains a window-based multi-head self-attention (W-MSA) module and a shifted window-based multi-head self-attention (SW-MSA) module, respectively. The "decoder" path of the network employs several simple convolutional layers for upsampling and dimensionality reduction. In conventional U-Net network, a skip connection is used to transfer the original spatial information from the encoder to a decoder with the same dimensions. However, the spatial information with speckle features in the encoder is often redundant in image reconstruction tasks. Using a skip connection to incorporate excess original spatial information into the decoder can potentially reduce the transmission efficiency of effective information between different dimensions. Here, we used a skip-layer excitation (SLE) module to establish a shortcut gradient flow between two dimensions in the decoder to transfer feature information between the two dimensions. The schematic diagram of the SLE module is depicted in Fig. 5**c**. The high-dimensional input feature map was applied to the adaptive average-pooling to downsample the sample to 4×4, and two convolutional layers were then used to obtain a one-dimensional vector. The length of the vector could be controlled by the output dimension of the convolution layer, which should be equal to the dimension of the low-dimensional feature map, so that the vector could directly be multiplied by the feature map. Therefore, through the SLE module, high-dimensional feature information could be transformed into low-dimensional spatial features.

When extracting feature information into high dimensions, the Euclidean distance between distinct data types in the feature space is larger than that at lower dimensions. Therefore, when random mode coupling occurs, distinct data points in low-dimensional feature spaces can be similar across different data types, making it difficult for conventional neural network to distinguish between two different data types in the low dimensional feature space, resulting in a lower accuracy in the neural network reconstruction. In contrast, the SLE module has large space distances between distinct data types in high dimensional feature spaces, making different data types less prone to crosstalk even if random mode coupling occurs. As a result, feature information in high-dimensional feature spaces can be more robust. In our FSUT-NN, high-dimensional feature information was transferred to a low-dimensional spatial feature using SLE modules with a robust gradient flow to compensate the disturbance caused by random mode coupling, thereby improving the generalization performance of the network [32].

## Image reconstruction

The optical setup of the imaging system is shown in Fig. 6. The light source used in this study was a narrowband laser with an output wavelength of 1550 nm. The emitted laser was subsequently converted into a Gaussian beam by a collimator lens, and the polarization direction was modulated using a quarter-wave plate and a polarizer (P1) prior to illumination with a spatial light modulator (SLM) to display image patterns. The beam carrying pattern information passed through a second polarizer (P2) to generate intensity-modulated patterns in the spatial domain (see Supplementary Note 1 for details). When employing the RCF as the transmission fiber, a $4f$ filtering system incorporating SPP was placed in front of the fiber to convert the beam into an OAM beam. The inset image in Fig. 6 is a scanning electron microscope image of the cross-section of the RCF, where the brightly illuminated ring indicates the high-refractive-index ring core. The modulated beam was then coupled into a 5-m-long RCF or MMF using a microscope objective and output at the distal end of the RCF or MMF through another microscope objective. The resulting output speckle was focused using a lens and recorded using a charge-coupled device (CCD) camera.

**Fig. 6: Experimental layout.** Optical setup. A schematic illustration of the experimental setup incorporating a SLM for image loading and transmission through the RCF or MMF. The inset image shows the cross-section of the RCF. QWP, quarter wave plate; P, polarizer; SLM, spatial light modulator; L, lens; SPP, spiral phase plate; FC, fiber coupler; CCD, charge-coupled device.

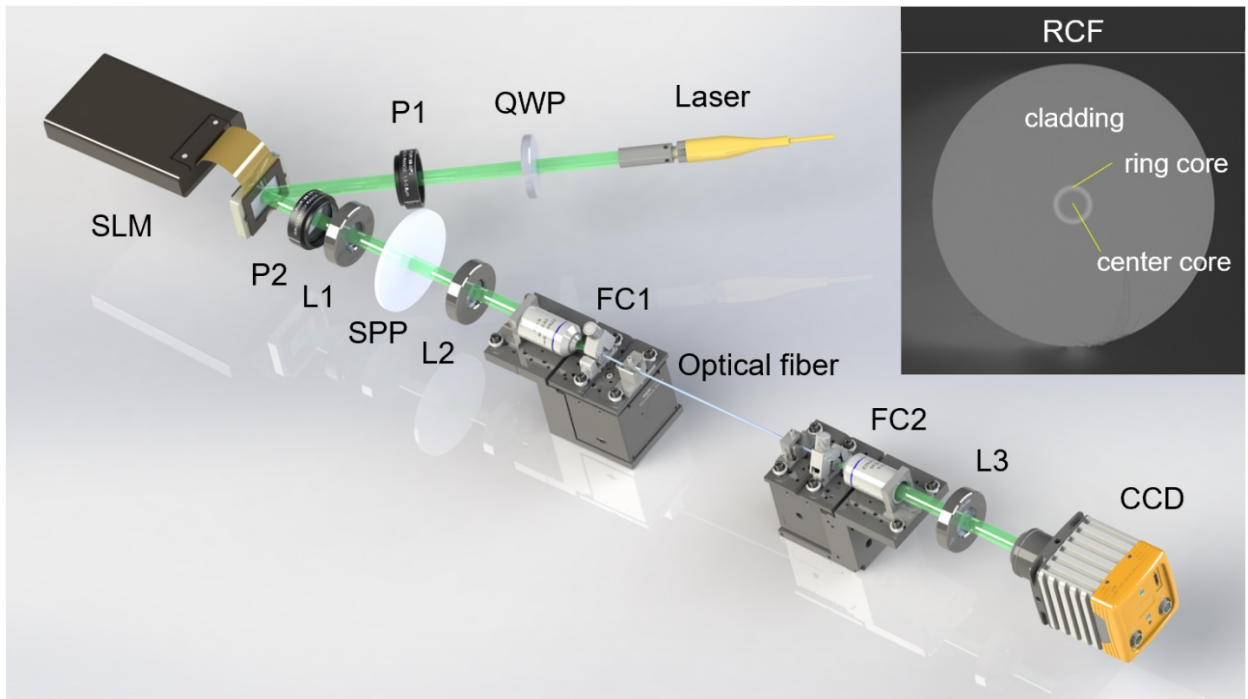

First, an experimental demonstration of the RCF-based imaging system was performed with a Modified National Institute of Standards and Technology (MNIST) dataset. A pattern of 10,000 handwritten digits was upsampled to 320×320 pixels using a bicubic interpolation method for display on the SLM. The corresponding speckle pattern collected with the CCD camera was downsampled from 320×256 to 128×128 pixels. To train the FSUT-NN, the obtained 10,000 image pairs were randomly divided into 9,500 training sets and 500 testing sets. After training for 100 epochs, the model was used to recover original images from the testing sets. The recovery results showed a high fidelity to the original image, with an average Pearson correlation coefficient (PCC) of 0.81. Some recovery results are shown in Fig. 7**a**. Furthermore, we replicated the image reconstruction process in a conventional MMF imaging system and also obtained a relatively good fidelity result with an average PCC value of approximately 0.80. Some of the recovery results with MMF are also shown in Fig. 7**b**. Notably, it is apparent that for image reconstruction tasks involving relatively simple edge information, both RCF-based and MMF-based imaging systems can transmit images with a similar performance.

**Fig. 7: Imaging of handwritten digit patterns. a** The input images of digits 0-9 selected from the test dataset were transformed into speckle images following their transmission in the RCF. Then, the images were reconstructed from the speckle images using our FSUT-NN. **b** The same input images was reconstructed with a conventional MMF imaging system.

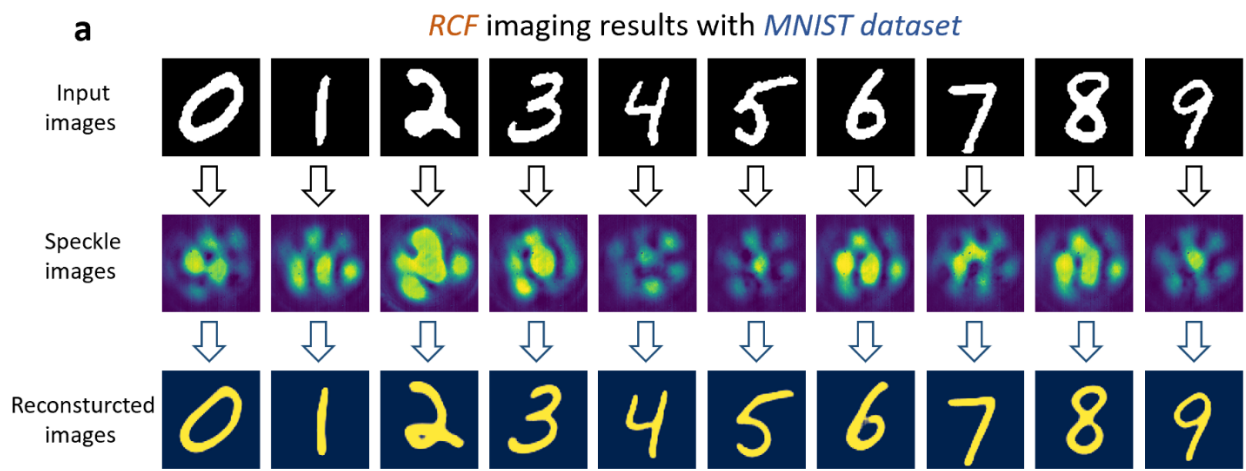

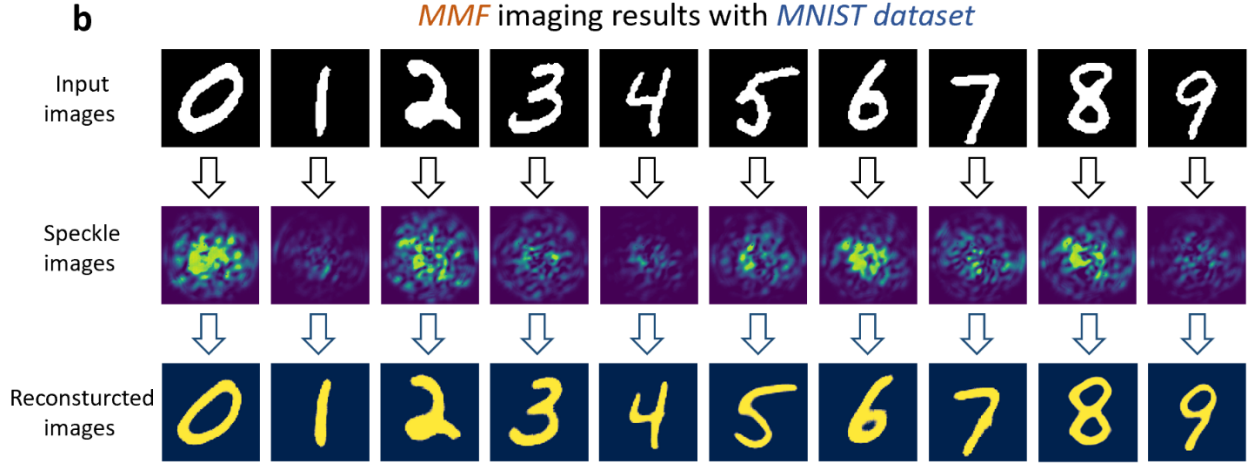

To verify the robustness of the RCF-based imaging system, images were transmitted through both the RCF and the MMF system under dynamic perturbation and static bending. The vibration amplitude and frequency could be adjusted by perturbing the fiber with an oscillator (see Supplementary Note 2 for details). First, based on the RCF imaging system, we collected the recovery images of the same test set under different vibration amplitudes varying from 0.5 mm to 3.5 mm at a fixed vibration frequency of 40 Hz, and we repeated the vibration experiment seven times at different locations within the optical fiber, with space intervals of 10 cm. The average PCC results of each group are shown in Fig. 8**a**, which also exhibits the reconstructed images of the number "8" at the initial location for each vibration amplitude. We also conducted the same vibration experiments in the MMF imaging system, and the corresponding results and reconstructed images of the number "8" are illustrated in Fig. 8**b**. Experimental results at different locations showed that as the vibration amplitude increased, the quality of the image reconstructed using the MMF system degraded significantly compared to those reconstructed using the RCF system. As the vibration amplitude increased from 0.5 mm to 3.5 mm, the average PCC of the MMF based image transmission decreased by about 0.121, while that of the RCF-based image transmission only decreased by around 0.055, highlighting the excellent robustness of the RCF based imaging system to dynamic environmental perturbation. To obtain the static stability, the transmission fiber was bent into different diameters. Specifically, we recorded five groups of images separately from the MMF and RCF imaging systems, respectively, with bend diameters ranging from 24 cm to 8 cm. The average PCC values of each testing set are shown in Fig. 8c, alongside the reconstruction result of the number "8" in these sets. Fig. 8d shows the comparison of some recovery results between the MMF and RCF imaging systems for an 8 cm fiber bending diameter. The average PCC of the MMF-based image transmission decreased to approximately 0.61, while that of the RCF-based image transmission remained at around 0.72. Therefore, the RCF-based imaging scheme exhibited an excellent robustness in the case of static bending, compared to that of MMF-based image scheme.

**Fig. 8: Robustness analysis. a** The average PCC values of the recovery results of the imaging system at different locations RCF vary with an increase in the disturbance amplitude. Some typical results are shown in the attached figures. **b** The average PCC values of the recovery results of the imaging system based on a MMF at different locations vary with an increase in the disturbance amplitude. Some typical results are shown in the attached figures. **c** The average PCC values of the recovery results of the imaging system based on RCF and MMF vary with an increase in the bending diameter. Some typical results are shown in the attached figures. **d** Comparison of some recovery results between the MMF and RCF imaging systems for a fiber bending diameter of 8 cm.

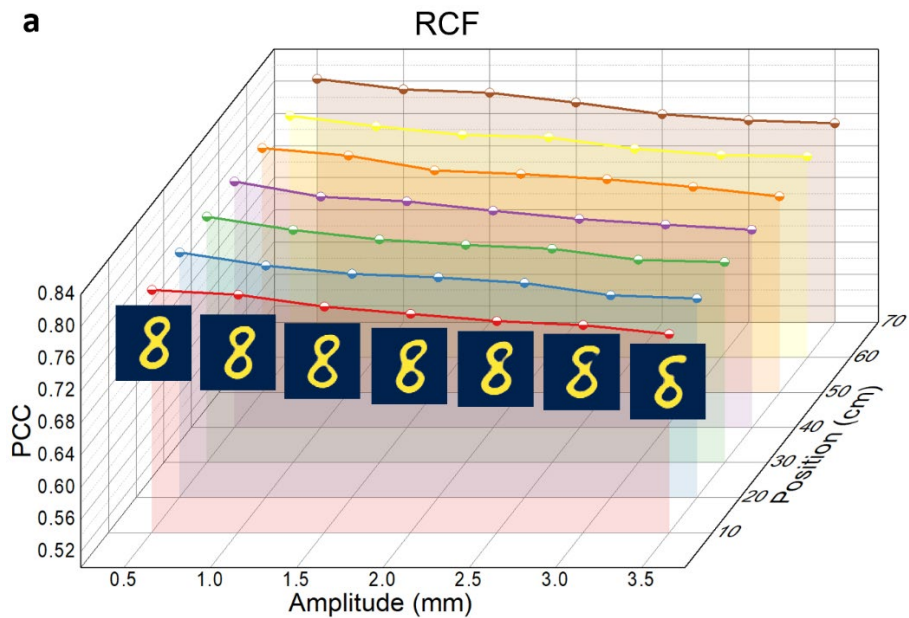

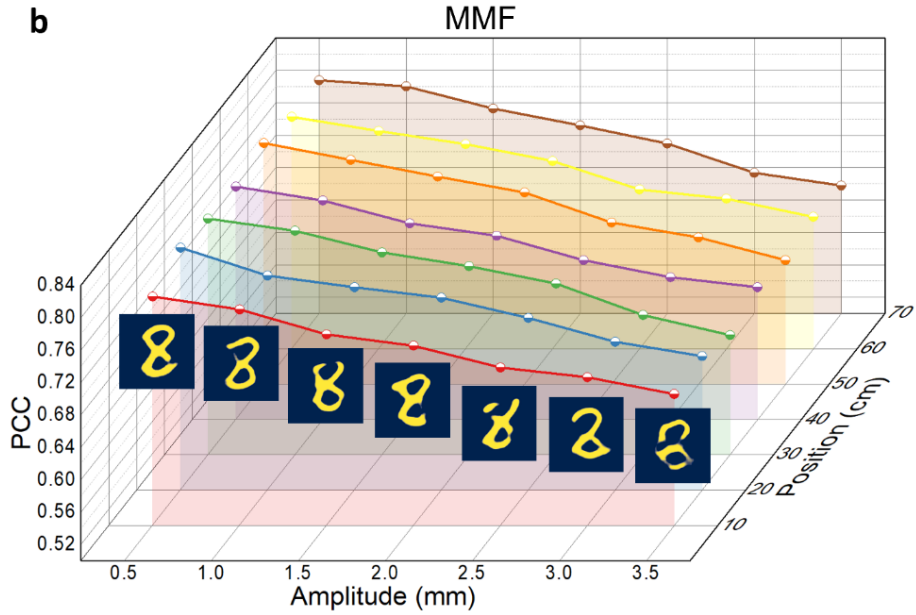

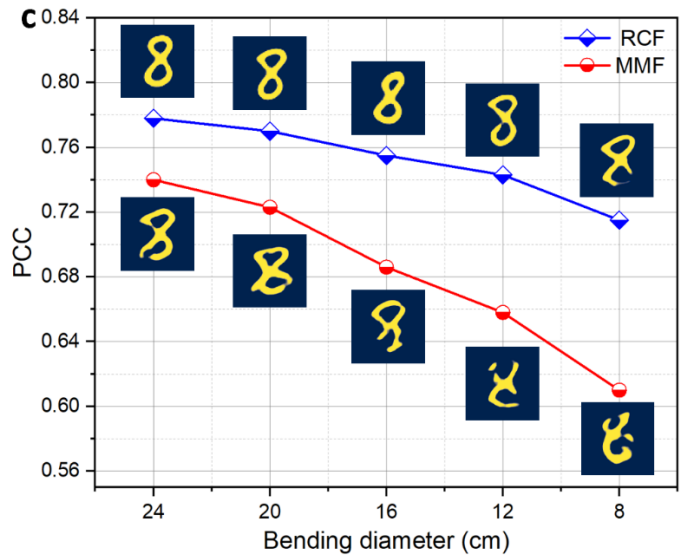

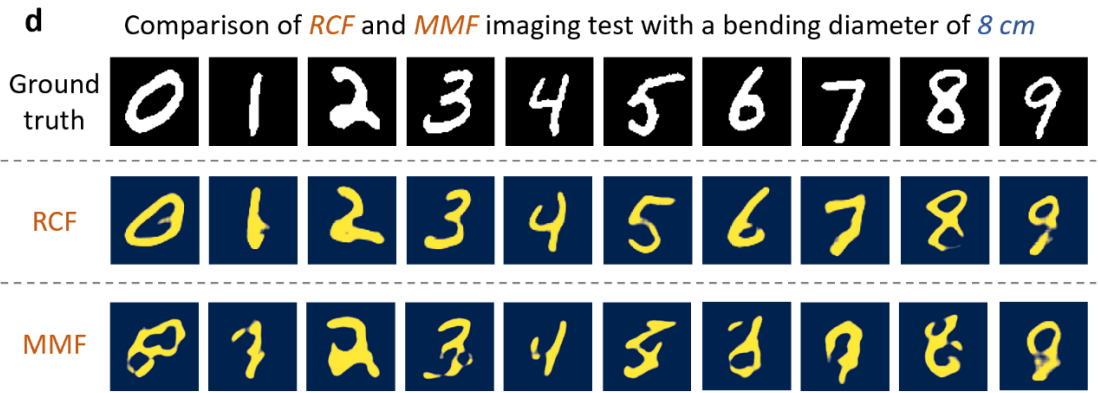

Subsequently, we evaluated the reconstruction performance of the RCF-based imaging scheme for images with complex structures and compared it with that of the conventional MMF-based imaging scheme. We first conducted the training and testing processes using 10,000 handwritten letters from the EMNIST dataset, as shown in Fig. 9**a**. The RCF-based imaging results yielded an average PCC value of 0.78 across all datasets, while the MMF-based imaging results had an average PCC value of 0.75. Then, we conducted the training and testing processes using 10,000 hand-painted patterns from the Quickdraw dataset, in which the images had a more complex structure, such as the envelope, star, bicycle, and umbrella. For similar imaging results, the RCF-based imaging system obtained an average PCC value of 0.77, which is higher than the value of 0.73 obtained by the MMF-based imaging system, as shown in Fig. 9**b**. The results showed that for image reconstruction tasks involving relatively more complex and diverse edge information, the RCF-based imaging scheme provides a better reconstruction fidelity than the MMF-based imaging scheme.

**Fig. 9: Imaging of other types of patterns. a** Several recovered images of handwritten letters reconstructed by the RCF and MMF imaging system. **b** Several recovered images of hand-painted patterns reconstructed by the RCF and MMF imaging systems.

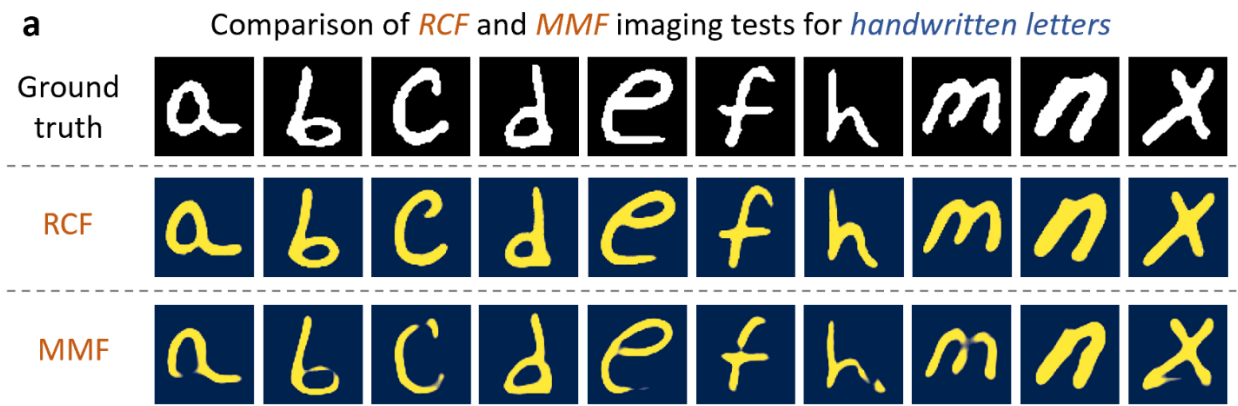

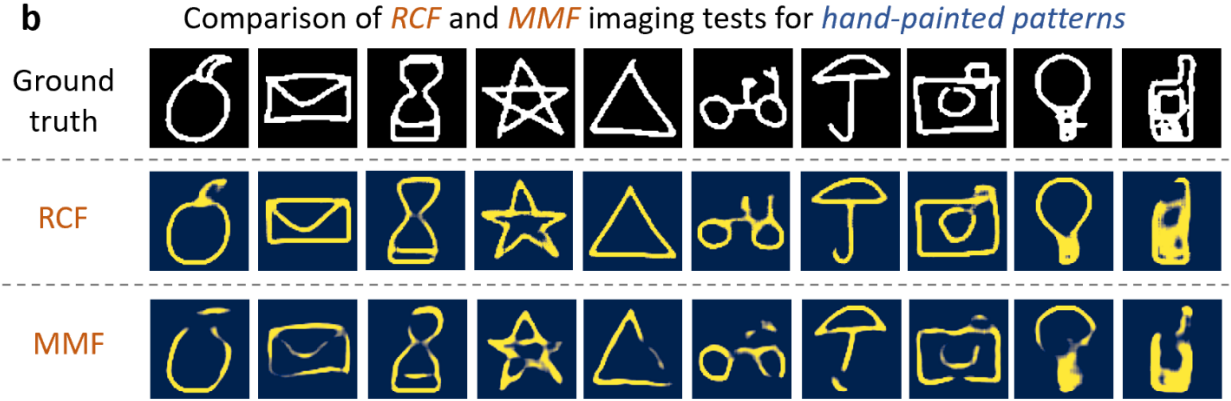

In previous experiments, the datasets for both training and testing images were of the same type. To evaluate the generalization ability of the FSUT-NN, we attempted to recover different types of images with the same learning and testing process. Generally, it is relatively difficult to reconstruct images using different learning and testing datasets. To this end, we generated an additional 20,000 training images based on the original images obtained from the Quickdraw dataset by cutting, shifting, and rotating the original patterns. We trained the proposed FSUT-NN and the conventional Swin-Unet model using the aforementioned datasets to reconstruct images of unlearned types from the Quickdraw dataset. Some exemplary results are shown in Fig. 10. The results show that compared with using a conventional Swin-Unet model, the predicted images recovered by the proposed FSUT-NN had higher accuracy, exhibiting an excellent fitting capability and generalization ability in unlearned image reconstruction tasks.

**Fig. 10: Imaging of unlearned patterns.** Several recovered images of new types that were not learned in the training stage recovered by the conventional Swin-Unet network and our FSUT-NN.

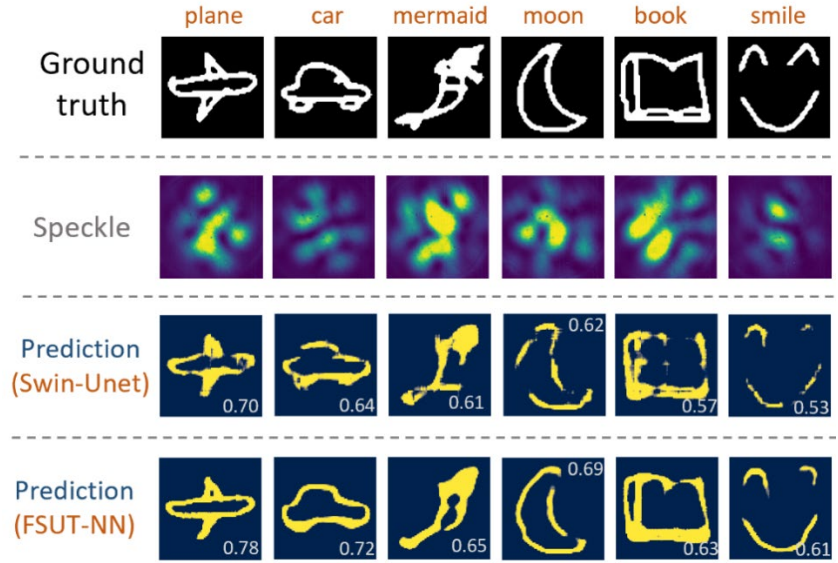

To evaluate the spatial resolution performance of the RCF-based imaging system with an OAM beam, several images from the 1951 USAF resolution test chart were reconstructed, which contained three white stripes with different spacings and thicknesses arranged horizontally and vertically. Since the imaging scale was the size of the image displayed on the SLM, we used a smaller area on the SLM to display the image and explore the minimum spatial resolution. In this experiment, the area of the images displayed on the SLM consisted of 32×32 pixels with a pixel size of 8 μm, and thus, each image had a spatial size of 256×256 μm$^2$. For comparative analysis, we conducted similar training processes in the MMF-based Gaussian beam imaging system. The results are compared in Fig. 11. The width of the horizonor or vertical white stripes increased, while the distance between them decreased accordingly. The edge resolution of the OAM beam was significantly better than that of the Gaussian beam. When the stripe width was narrow, the edge of the stripe became blurred or could not even be reconstructed using a Gaussian beam, while the edge of the stripe maintained a sharp shape with the OAM beam. As the stripe width increased and the spacing decreased, the stripe edges imaged by the Gaussian beam merged together, while the OAM beam imaging could still effectively separate the stripe edges with correlated shapes with a resolution of up to 16 μm. Therefore, the OAM beam imaging with RCF transmission achieved a high image resolution with an edge enhancement.

**Fig. 11: Edge enhancement performance.** Recovery results of images similar to the USAF 1951 obtained through MMF imaging system with Gaussian light and RCF imaging system with OAM light.

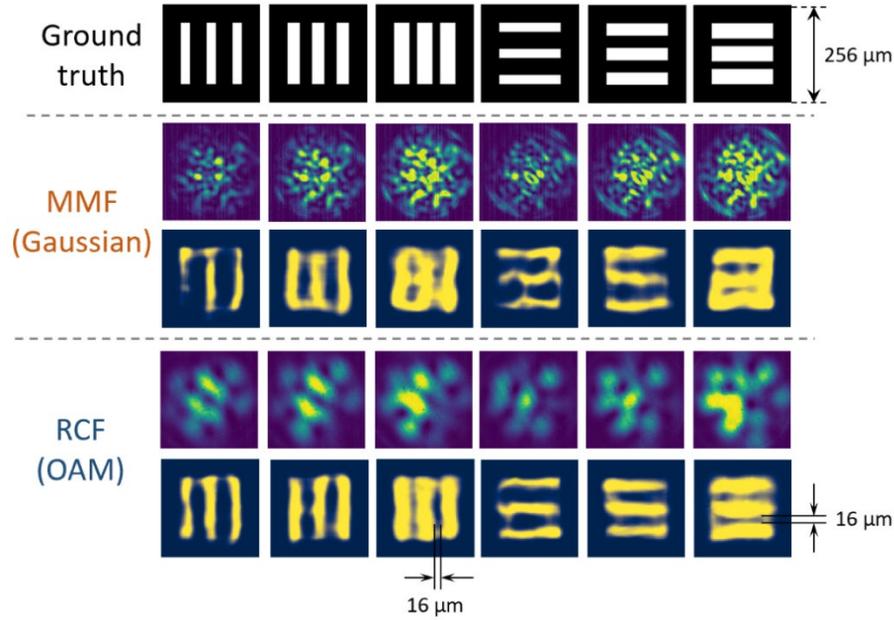

Finally, we verified the feasibility of imaging biological tissues using the proposed scheme. We collected cells and tissues from animal and plant tissues to produce four types of biological slices: onion epidermal scale cells, HeLa cells, gastric gland tissues, and myocardial tissues. As depicted in Fig. 12**a**, we designed a dual-channel system to image the prepared biological slices with a tungsten-halogen lamp and a 1550 nm laser as illumination sources. Specifically, the image captured by the camera was used as the ground truth. On the other path, the light beam emitted from the laser source was converted into an OAM light beam and coupled into the optical fiber after imaging the biological slices. Following optical fiber transmission, the output speckle patterns were captured by the CCD camera and combined with the ground truth as a training image. To achieve an excellent image resolution, we downsampled the speckle pattern to 256×256 pixels, resulting in a reconstructed image size of 256×256 pixels. Subsequently, the speckles in the testing set were reconstructed using the trained FSUT-NN. We also carried out comparative experiments based on the MMF and RCF imaging systems. To ensure the consistency of the training and testing image sets in the MMF and RCF-based imaging processes, we carefully controlled the imaging position of the biological slices with the aid of a motorized moving platform. The recovery results are shown in Fig. 12**b**. These show that the PCC values of the four recovery biological slice images using the RCF-based imaging system were 0.93, 0.86, 0.71, and 0.60, respectively, which were higher than those of 0.87, 0.82, 0.62, and 0.45 obtained with the conventional MMF-based imaging system. For the onion epidermal scale cells and HeLa cells, both MMF and RCF-based imaging systems exhibited a high reconstruction fidelity with slightly biased PCC values due to the simple image background. However, for gastric adenocarcinoma tissue and myocardial tissue which contained complex image details, the MMF-based imaging results showed a poor reconstruction fidelity, while the RCF-based imaging system still had a relatively high PCC value and good detail features. Therefore, the proposed RCF-based imaging system had an ultrahigh resolution compared to that of the conventional MMF-based imaging system.

**Fig. 12: Reconstruction of biological image. a** Configuration of the optical imaging system of the biological sections and the imaging observation of gastric gland tissues (the stomach image by [https://www.scientificanimations.com] titled [3D Medical Animation Stomach Structure] was used under CC-BY-SA-4.0 license, available at [https://www.scientificanimations.com/wiki-images/].) **b** Imaging and prediction results of the onion epidermal scale cells, HeLa cells, and myocardial tissues; the PCC values of each reconstruction are displayed in yellow in the bottom right corner.

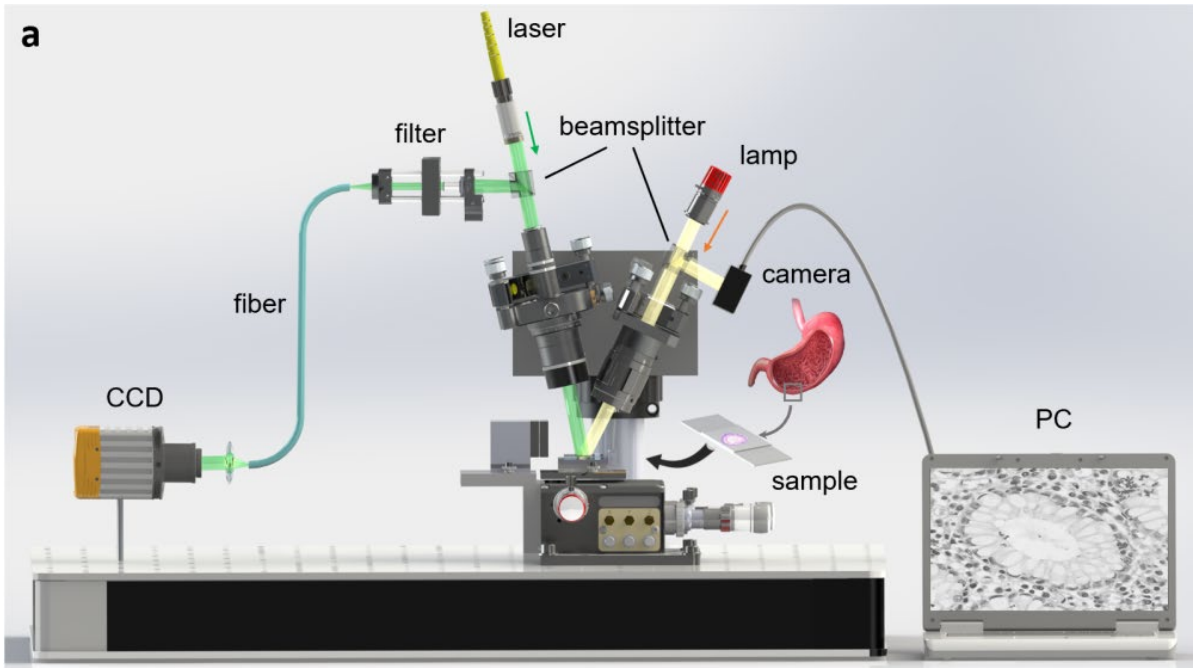

**Conclusion**

In conclusion, we have proposed and experimentally demonstrated a robust super-resolution imaging method based on a RCF with OAM. The RCF-OAM imaging system overcomes the challenges of dynamic perturbations and low imaging resolution commonly faced by MMF imaging technologies. By utilizing weakly-coupled inter-MGs and strong-coupled intra-MGs of OAM modes, the RCF-OAM system provides robustness to environmental disturbances, ensuring reliable image transmission. Additionally, the integration of a SPP as a vortex filter enhances edge sharpness, improving image resolution. The proposed FSUT-NN enhances the resilience of the developed RCF-OAM imaging system against environmental perturbations, further improving system stability. Our results suggest that this imaging scheme provides better robustness and higher resolution compared to MMF imaging scheme. Moreover, we validated the excellent performance of the RCF based imaging scheme in the imaging of biological slices, demonstrating the practicality of our scheme. Our pioneering RCF-OAM imaging system providing an effective and low-cost imaging method, which shows the potential broad applications in biological imaging and industrial non-destructive testing, paving the way for future advancements in microendoscopy.

## Methods

### Experiments

The laser source was a narrowband laser with an output wavelength of 1550 nm. The SLM (PLUTO-2.1 Phase only, Holoeye) consisted of 1920×1080 microdisplays, each being 8 μm in size. The inner/outer radius of the core of the RCF was 4 μm/8.5 μm, and the core and cladding diameters of GI-MMF were 50 μm/125 μm. The refractive index differences of the RCF and GI-MMF between the fiber core and cladding were both around 1%. The CCD camera (Bobcat 320, Xeneth) utilized an in-house developed, temperature-stabilised InGaAs detector with a 320×256 pixel resolution, offering frame rates of either 100 Hz or 400 Hz.

In the process of collecting speckles, a computer program was used to autonomously control the image loading on the SLM and the capture of speckles on the CCD camera. The collected speckle data and ground truth image were upsampled or downsampled through bicubic interpolation to yield a uniform size, which was input into the neural network during the training process. The FSUT-NN model was trained for a maximum of 100 epochs, with a batch size of 200, employing the Adam optimizer with a learning rate of $1\times10^{-4}$ to minimize the loss function. For sparse image reconstruction, the binary cross-entropy (BCE) was utilized as the loss function. For the greyscale images, where pixel values were not binary in nature, the BCE loss function was inapplicable, and instead, the mean squared error was used as the loss function. The network was implemented using the PyTorch 1.10 software, running on an NVIDIA GeForce GTX 3090 graphics processing unit.

## Supporting Information

Supporting Information is available from the Wiley Online Library or from the author.

## Acknowledgements

This work was supported by the National Key R&D Program of China from Ministry of Science and Technology (2019YFA0706300); National Natural Science Foundation of China under Grants (62105026, 62205023); The Fundamental Research Funds for the Central Universities; Beijing Municipal Natural Science Foundation (4222075).

## Conflict of Interest

The authors declare no competing interests.

## Data Availability Statement

The data that support the findings of this study are available from the corresponding author upon reasonable request, and the code for few-shot U-Transformer neural network and analysis code can be found at https://github.com/HEMessiah/few-shot-U-Transformer-neural-network.